\documentclass[aps,prc,preprint,amssymb,amsmath,superscriptaddress,floatfix,10pt]{revtex4-1}
\usepackage{mathptmx}
\usepackage{graphicx}
\usepackage{color}  
\makeindex
\newcommand{\textblue}{\textcolor[rgb]{0.00,0.07,1.00}}

\begin{document}

\title{Location and symmetry of superconductivity in neutron stars}
\author{Dmitry Kobyakov}
\date{2023-10-25}
\email{dmitry.kobyakov@appl.sci-nnov.ru}
\affiliation{Institute of Applied Physics of the Russian Academy of Sciences, 603950 Nizhny Novgorod, Russia}

\begin{abstract}
Earlier, it was a standard assumption that the entire core of neutron stars is superconducting. 
However, the matter contents in the inner core has been unknown even qualitatively, because the density of matter in that region is expected to be higher than the nuclear saturation density 0.16 $\mathrm{fm}^{-3}$.
As a consequence, no reliable model exists that would describe the neutron star matter in the inner core of neutron stars. 
Thus, a possibility of presence of normal, nonsuperconducting, plasma in the inner core cannot be excluded as of today. 
This point is supported by the numerical calculations performed in \cite{Kobyakov2023b}. 
The calculations are based on the equation of state and the proton Cooper pairing gap energy derived from the chiral effective field theory. 
The numerical results show that the superconducting gap goes to zero beyond the depth about 1 km below the crust-core boundary. 
Given that the stellar radius is of the order of 12 km, therefore the superconducting proton matter is expected to exist only in a thin layer at the tip of the outer core. 
Recently it has been realized that the symmetry of superconductor is anisotropic in the lasagna region of the pasta phases located at the bottom of the crust. 
However the question of whether this symmetry is continuous or discreet was unsolved. 
The numerical calculations performed in \cite{Kobyakov2023b} have shown that the tunneling rate between the adjacent slabs in the entire range of the corresponding densities is negligibly small. 
Thus, a discreet model is necessary for the description of the lasagna region. 
Uncertainties and future directions of the research are discussed.
\end{abstract}

\maketitle

The location of superconducting protons and symmetry properties of the order parameter are crucial for the spectrum of hydromagnetic waves in neutron stars.
The hydromagnetic waves transfer energy from the inner part of the star to its outer layers in the observable processes such as the magnetar giant flares and the following quasiperiodic oscillations, the glitches of the spin frequency.
Thus, understanding of the spectrum of the hydromagnetic waves is a crucial part of astrophysical models designed for generating novel knowledge about the structure of the neutron star matter.

The existence of superconductivity in neutron star has been for the first time considered in 1969 in \cite{BPP1969a} and it has been shown that the strong proton-proton interaction must lead to the S-wave Cooper pairing and to superconductivity in the core (where the nuclear matter is expected to be a uniform quantum liquid).
Since then, the standard physical picture has been that the neutron star core is completely filled by the proton superconductor with the isotropic order parameter.

However, recent calculations \cite{Kobyakov2023b} based on the energy-dependent proton Cooper pairing gap energy derived from the chiral effective field theory of baryons \cite{LimHolt2021}, have shown that the superconductor fills only a thin layer at the tip of the core with thickness of the order of 1 kilometer. 
To the best of my knowledge this result has not been discussed before and therefore is novel.

The conclusion that the superconductor fills only a thin layer of the core does not depend neither on the equation of state (EoS) in the crust, nor on the polytropic exponentials in the inner core (where the matter density is higher than the nuclear saturation density).
This result is shown in figure 11 of \cite{Kobyakov2023b} and has been obtained from the solution of the equations of the force balance between the gravitational force and the pressure gradient supplemented by the EoS consisting of three parts:
The first is the EoS of the solid crust, the second is the EoS of the outer core and the third is the extrapolation of the EoS into the inner core.

\begin{figure}
\includegraphics[width=3.5in]{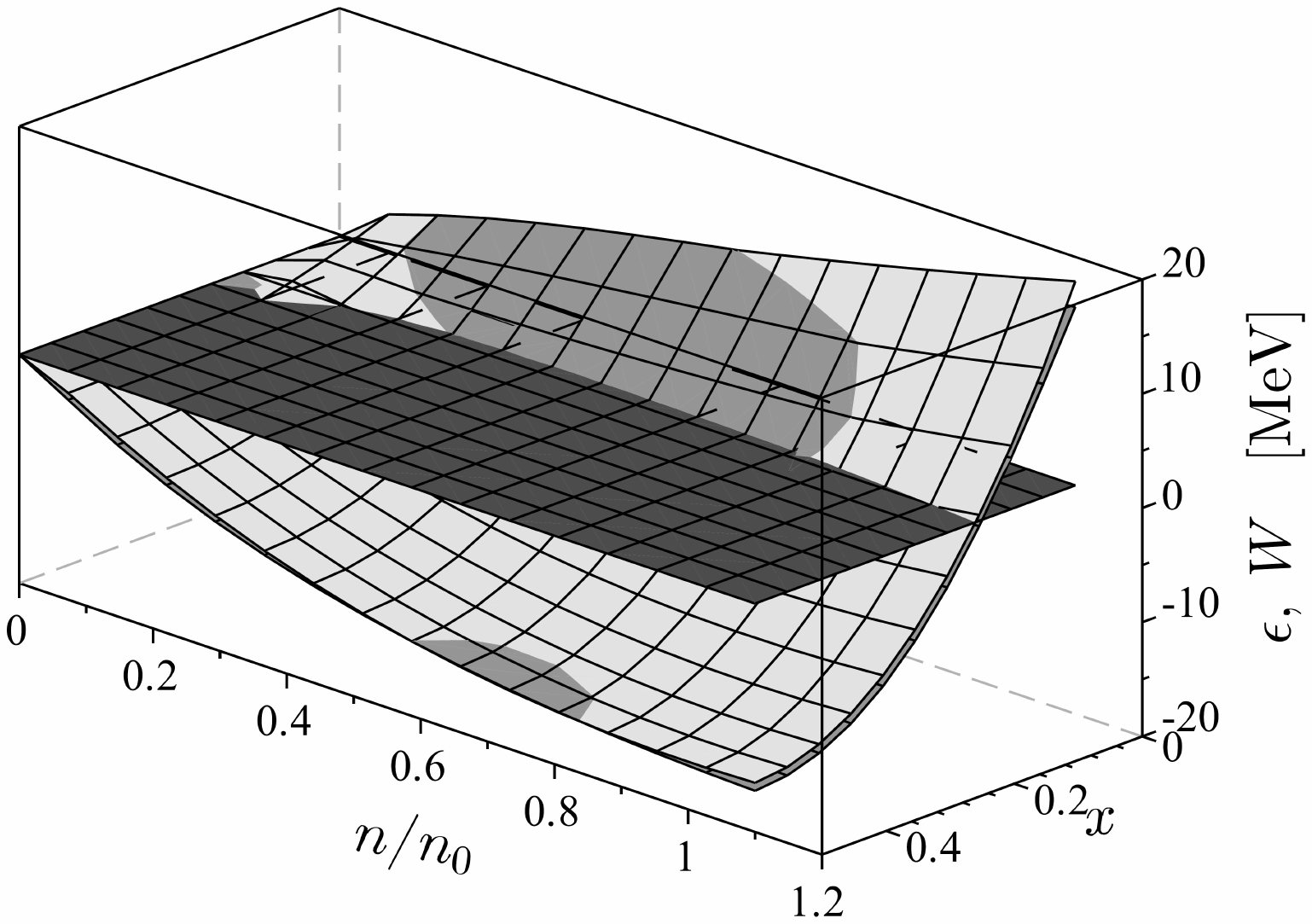}
\caption{}
\end{figure}
The EoS of the crust has been taken from the literature with the data obtained from the Barcelona-Catania-Paris-Madrid EoS (see the references in \cite{Kobyakov2023b}).
However, for the self-consistency it is desirable to calculate the pressure from the same EoS as used in the outer core.
In the outer core, I used the EoS derived from the chiral effective field theory.
The energy per baryon is given by the following expression:
\begin{equation}
 \label{e_x_n} \varepsilon(n,x)=\varepsilon_0\left[\frac{3}{5}\left[x^{\frac{5}{3}} + \left(1-x\right)^{\frac{5}{3}}\right]\left(\frac{2n}{n_0}\right)^{\frac{2}{3}}-\left[\alpha_1\left(x-x^2\right) + \alpha_2\right]\frac{n}{n_0}+\left[\eta_1\left(x-x^2\right) + \eta_2\right]\left(\frac{n}{n_0}\right)^\gamma \right].
\end{equation}
where $\varepsilon_0=36.84$ MeV, $\alpha_1=2\alpha-4\alpha_L$, $\alpha_2=\alpha_L$, $\eta_1=2\eta-4\eta_L$, $\eta_2=\eta_L$.
The function $\varepsilon(n,x)$ is shown in Fig. 1 by light-gray. 
For comparison, in Fig. 1 the energy per baryon from the work by Baym, Bethe and Pethick (1971) (see the references in \cite{Kobyakov2023b}) by middle-gray; for convenience, the dark-gray shows the zero surface.
The mass density includes the rest mass of protons and neutrons, their interaction energy and the relativistic mass of the electrons:
\begin{equation}\label{EoSOuterCore}
  \rho=m_pxn+m_n(1-x)n+\frac{n\varepsilon}{c^2}+\frac{(9\pi)^{2/3}}{4}\frac{\hbar}{c}\left(xn\right)^{\frac{4}{3}},
\end{equation}
where $x={n_p}/{n}$.
From the empirical properties of nuclear matter we have the (minus) binding energy per baryon $\omega_0=16\;{\rm MeV}$ and the pressure of atomic nucleus $P_{\rm nuc}(n=n_0,x=1/2)=0$.
From these properties I obtain
\begin{equation}\label{parameters}
\alpha=\frac{4}{5} + \frac{2\gamma}{\gamma-1}\left(\frac{1}{5} + \frac{\omega_0}{\varepsilon_0}\right), \quad \eta=\frac{2}{\gamma-1}\left(\frac{1}{5} + \frac{\omega_0}{\varepsilon_0}\right).
\end{equation}
The incompressibility, symmetry energy and its slope are given by
\begin{eqnarray}
\label{K} &&   K=9\varepsilon_0\left[-\frac{2}{15} + \gamma\left(\frac{1}{5} + \frac{\omega_0}{\varepsilon_0}\right)\right], \\
\label{S0} &&   S_0=\varepsilon_0\left( \frac{3}{5}2^{2/3} + \frac{\omega_0}{\varepsilon_0} - \alpha_L + \eta_L \right), \\
\label{L} &&   L=3\varepsilon_0\left( \frac{2}{5} - \alpha_L + \gamma\eta_L \right).
\end{eqnarray}
In the numerical calculations in \cite{Kobyakov2023b}, I used two sets of the parameters: $(\gamma,\alpha_L,\eta_L)=(4/3,1.385,0.875)$ and $(\gamma,\alpha_L,\eta_L)=(1.45,1.59,1.11)$, which lead, correspondingly, to
\begin{eqnarray}
\label{p1} && (K,S_0,L)=(236\,{\rm MeV},32.3\,{\rm MeV},20.1\,{\rm MeV}), \\
\label{p2} && (K,S_0,L)=(261\,{\rm MeV},33.4\,{\rm MeV},46.4\,{\rm MeV}).
\end{eqnarray}
In the inner core, I used the generalized polytropic EoS, where the pressure was defined in three consequent regions of the matter density as following:
\begin{equation}\label{EoSinnerCore}
  P[\rho(r)]\propto\rho^\Gamma,
\end{equation}
where $r$ is the radial coordinate and the values of $\Gamma$ are given in the caption to figure 2 in \cite{Kobyakov2023b}.

From the other hand, at densities higher than the nuclear saturation density the contents of the neutron star matter is even qualitatively unknown and a reliable description of the structure of the inner core does not exist.
Therefore, at present it is impossible to exclude the existence of the normal plasma (without superconductivity) in the inner core of neutron stars.

In \cite{Kobyakov2023b}, I have considered the structure of the superconductor at the crust-core boundary.
In 2018, it has been for the first time noticed \cite{Kobyakov2018,KobyakovPethick2018} that the ground state of neutron star matter at the crust core boundary corresponds to the structure when the proton liquid is distributed over thin slabs, thus making the symmetry of the order parameter anisotropic.
Thus, a question arises whether the anisotropic symmetry is continuous or discreet.
In case of a continuous symmetry, the superconducting density may be described by a tensor.
On the contrary, if the superconductor is distributed over the slabs which are not connected by the Josephson tunneling, the system should be described by a discreet model.
In order to specify the symmetry type in \cite{Kobyakov2023b}, I have calculated probabilities of quantum mechanical tunneling between the adjacent layers (in figures 6-9) and found that in mostly, the tunneling that defines the Josephson coupling is negligibly small.
It then follows that the ground state of the superconductor should be described by the discreet model.

The results obtained in \cite{Kobyakov2023b} specify the basic assumptions of future models for the calculation of the magnetic torque exerted on the structure by the stellar magnetic induction.
Such calculations are necessary for investigation of the role of thermal fluctuations in the neutron star matter in order to find out whether the pasta region is ordered or disordered at typical realistic conditions.

In the future work, it is necessary to investigate the following issues. (i) Specify the radial position and independently confirm the existence of the lasagna region of the pasta structure within other EoS. (ii) Investigate whether the ground state in the realistic neutron star matter is ordered or disordered. (iii) Specify the thickness and separation of the slabs. (iv) Investigate the uncertainty  related to the mutual orientation of the slabs and the magnetic field. (v) Study the hydromagnetic waves at the boundary between the superconducting and the normal plasmas.

\emph{Acknowledgements.} This work was supported by the Center of Excellence ``Center of Photonics'' funded by The Ministry of Science and Higher Education of the Russian Federation, contract No. 075-15-2022-316.

\emph{Translated by the author.}

\end{document}